
\magnification=\magstep1


\newif\iffigs\figsfalse

\input amstex.tex
\iffigs
  \input epsf
\else
  \message{No figures will be included. See TeX file for more
information.}
\fi
\documentstyle{amsppt}

\openup1.5\jot
\NoRunningHeads
\NoBlackBoxes
\topmatter
\line{\hfil RU-93-44}
\line{\hfil hep-th/9309145}
\medskip
\centerline{\bf SOME SPECULATIONS ABOUT BLACK HOLE}
\centerline{\bf ENTROPY IN STRING THEORY}
\bigskip
\centerline{\smc By Leonard Susskind\footnote{susskind\@dormouse.stanford.edu}}
\smallskip
\centerline{Physics Department}
\centerline{Stanford University}
\centerline{Stanford, CA 94305-4060}
\centerline{and}
\centerline{Department of Physics and Astronomy}
\centerline{Rutgers University}
\centerline{Piscataway, NJ 08855-0849}

\abstract{The classical Bekenstein entropy of a black hole is argued
to arise from configurations of strings with ends which are frozen on the
horizon.  Quantum corrections to this entropy are probably finite
unlike the case in quantum field theory.  Finally it is speculated
that all black holes are single string states.  The level density of
strings is of the right order of magnitude to reproduce the Bekenstein
entropy.}
\endabstract
\endtopmatter
\document
{\bf I.  Introduction}
\medskip
There are some puzzles concerning the entropy of black holes which I
would like to consider from the point of view of string theory.  First
of all the meaning of the Bekenstein entropy
$$
S_B = {1\over 4} \, {\text{Area}\over 4G\hbar}\tag 1.1
$$
has always been mysterious.  Entropy, as generally understood, has to
do with the counting of configurations of some set of degrees of
freedom.  What the degrees of freedom of the horizon are and why they
give entropy of order $1/\hbar$ has remained
obscure.\footnote{Hereafter $\hbar$ will be set to 1}
\medskip
The second puzzle concerns the higher order quantum corrections to the
entropy.  G 'tHooft has emphasized\cite{1} that conventional quantum
fields contribute an ultraviolet divergence to $S$ which blows up at
the horizon.  This is despite the fact that no curvature or other
invariant signal becomes large.  The ultraviolet divergent entropy is
proportional to the area and although down by a factor of $\hbar$ from
$S_B$ it is infinitely larger.
\medskip
It has not been properly appreciated except by 't Hooft
that this ultraviolet divergence of $S$ is the same problem as
Hawking's information paradox and that any theory which naturally
produces a finite entropy will also solve this problem.  From the
perspective of a distant observer nothing ever reaches the horizon.
Instead all matter settles into layers which eternally sink
toward it.  If the entropy of matter near the horizon is infinite,
indefinite amounts of information can be stored arbitrarily close by.
This information can not be emitted until the horizon shrinks to
quantum mechanical proportions and perhaps not even then.  By
contrast, a theory in which the information storage capacity is
finite has no choice but to reemit information as the horizon shrinks.
In a previous paper\cite{2} I showed that in string theory, from the
outside observers vantage point, the substance of infalling strings not
only never reaches the horizon but never entirely sinks past the
stretched horizon.  Therefore it seems
appropriate to ask whether string theory leads to a finite entropy.
\medskip
The last puzzle concerns the connection between the spectrum of black
holes and that of unperturbed strings.  In both cases that level
density increases rapidly with mass.  Furthermore, most of the
spectrum of strings must actually be black holes since they lie within
their Schwarzschild radii.  Nevertheless I do not know of any
speculation that the two spectra may really be the same.\cite{3} In fact
at first sight such a suggestion seems nonsensical.  The level density
of black holes grows like $\exp 4\pi M^2$ while that of strings is
exponential in the first power of the mass.  We shall see  that this
argument is wrong and that the two spectra, when properly interpreted,
could easily be the same.
\bigskip
{\bf II.  Thermodynamics in Rindler Space}
\medskip
I will begin by outlining the procedures involved in constructing the
thermodynamics of Rindler space.  The euclidean continuation is
ordinary flat space in cylindrical polar coordinates
$$
ds^2 = r^2 d\theta^2 +dr^2 + dx^2_\bot \tag2.1
$$
where the angular variable $\theta$ is the euclidean time, $r$ is the
radical variable (not to be confused with the Schwarzschild coordinate)
and $X_\bot$ is the 2 space parallel to the horizon.  The horizon
itself is the surface $r=0$.
\medskip
The Rindler hamiltonian $H_R$ is the generator of $\theta$-rotations.
$$
H_R = {\partial\over \partial\theta}\tag2.2
$$
\medskip
The partition function from which thermodynamics is derived is
$$Z = Tr \exp\{-\beta H_R\}\tag2.3
$$For physical applications $\beta$ should be set equal to 2$\pi$.
For the purposes of thermodynamic analysis it must be left as a free
variable.  Note that varying $\beta$ is equivalent to introducing a
conical singularity at the horizon with an angular deficit
$2\pi-\beta$.
\medskip
The free energy $F(\beta)$ is given by
$$
F = {1\over \beta} \log Z\tag 2.4
$$
and the entropy by
$$
S = {\partial F\over \partial T_R}\tag2.5
$$
where the Rindler temperature is $T_R = 1/\beta$.  Eq. 2.5 is the
reason we need to be able to vary $\beta$ away from $2\pi$.
\medskip
Consider first the thermodynamics of a free scalar field.  The
partition function can be carried out as a sum over first quantized
closed path particle trajectories in a well known manner.  The only
new ingredient is the conical singularity at $r = 0$.
\medskip
It is easy to see that paths which do not wind around $r=0$ contribute
no $\beta$ dependence to the free energy.  This means that they can be
dropped when calculating the entropy.  Thus we find that the entropy
is the first order variation with respect to $\beta$ of the sum of
paths which wind one or more times around the horizon.  This principle
extends to interacting field theories described in terms of networks
of paths forming Feynman diagrams.  Only those networks which
topologically encircle $r= 0$ contribute to entropy.  It is clear that
the entropy divergences found in conventional quantum field theory are
due to very small loops near $r = 0$.
\medskip
Now I want to study black hole entropy using an analogous method.
Consider the euclidean continuation of the Schwarzschild metric in
which time is periodic.  The geometry near $r = 0$ is identical to
Rindler space.  Far from the horizon 2.1 is modified to
$$
\align
ds ^2 &= 16 m^2 G^2 d\theta^2 + dr^2\\
 &+ \text{transverse metric}\\
 &= - dt^2 + dr^2 +\text{transverse}
\tag2.6
\endalign
$$
where
$$
t = - 4 \imath m G\theta\tag2.7
$$
is the Schwarzshild time coordinate.
\medskip
The ordinary energy (mass) of a black hole is conjugate to
Schwarzshild time and the conventional temperature is defined in terms
of this energy.  The energy and Schwarzshild temperature are given in
terms of the mass by
$$
\align
S &= 4\pi M^2 G\\
Ts &= {1\over 8\pi MG}
\tag2.8
\endalign
$$
\medskip
In transforming energy from Schwarzshild to Rindler coordinates care
must be taken to insure that the Rindler energy $ E_R$
is conjugate to $\imath\theta$.  We assume $M$ is conjugate to $t$.
Thus
$$
\align
[E_R(M), \imath\theta] &= \imath\\
[M, t] &= \imath
\tag2.9
\endalign
$$
\medskip
Using 2.7 and the usual properties of commitators gives
$$
E_R = 2M^2 G\tag2.10
$$
\medskip
It is interesting that both entropy and Rindler energy are extensive
functions of the horizon area ($A = 16\pi M^2 G^2$).
$$
\align
E_R &= {A\over 8\pi G} = 4\pi 2M^2G\\
S &= {A\over 4G} = 4\pi M^2G\tag 2.11
\endalign
$$
\medskip
To find the Rindler temperature we use the first law of thermodynamics
$$
dE_R = T_R dS\tag 2.12
$$
which gives
$$
T_R = {1\over 2\pi}\tag 2.13
$$
\medskip
To compute from first principles directly in Rindler coordinates we
must calculate the free energy of a euclidean black hole with an angle
deficit of ($2\pi - \beta$).  Again this introduces a conical
singularity at the horizon.  The lowest order contribution to $F$ is
order $1/\hbar$ and is given terms of the classical action.
$$
F = {1\over \beta} \, \text{Action}\tag 2.14
$$
The action consists of three distinct terms.  The first is
proportional to the integrated Ricci scalar which vanishes by virtue of
the Einstein equations.  The second involves the extrinsic curvature
at the boundary at large $r$.  This term is proportional to
$\beta$ and does not contribute to the entropy.  The third
contribution is due to the curvature delta function at the conical
singularity.  It is proportional to the horizon area and to
the deficit angle $2\pi- \beta$.  Explicit calculation gives
$$
F = {\beta- 2\pi\over \beta}  \, {\text{Area}\over 8\pi G}\tag2.15
$$
and
$$
S = {\partial F\over \partial T_R} = {\text{Area}\over 4G}\tag2.16
$$
\bigskip
{\bf III.  Black Hole Entropy in String Theory}
\medskip
Evaluating the partition function in a conical background provides a
general framework for calculating black hole entropy including the
classical Bekenstein term and quantum corrections from both matter and
gravitational field.  To apply it to string theory requires formulating
the theory in backgrounds with small but arbitrary angle deficits.
This has not been done. I will therefore restrict my remarks to certain
general features.  The free energy is given as a sum over world sheet
configurations of arbitrary genus.  By an argument similar to that
used in field theory, the only nonvanishing contributions to the
entropy come from world sheets which in some way wrap around or touch the
singularity at the horizon.  For example a torus can surround the
horizon as in fig(1)
\iffigs
\midinsert
\hfil\leavevmode\epsfysize=7pc \epsfbox{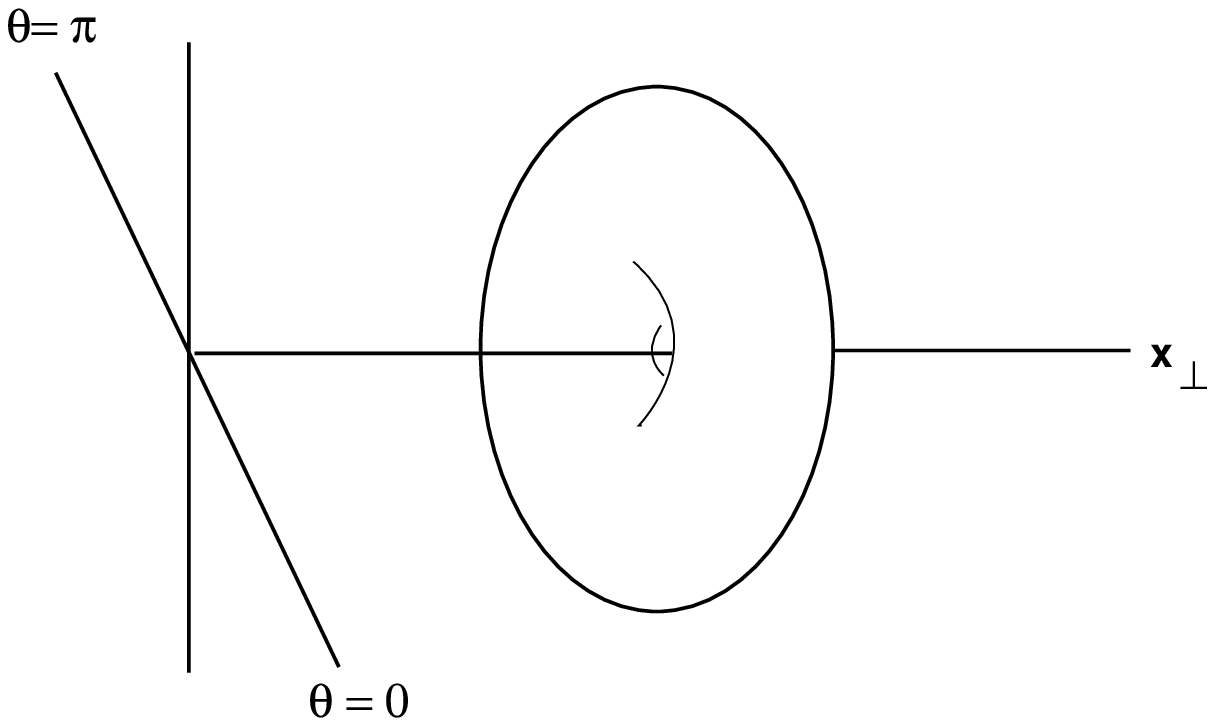}\hfil
\botcaption{Figure 1. }\endcaption
\endinsert
\medskip
\fi
This configuration describes the contribution to the entropy of a free
closed string.  This is seen by slicing the figure at some fixed
euclidean time such as $\theta = 0$  (See fig 2)
\iffigs
\midinsert
\hfil\leavevmode\epsfysize=7pc \epsfbox{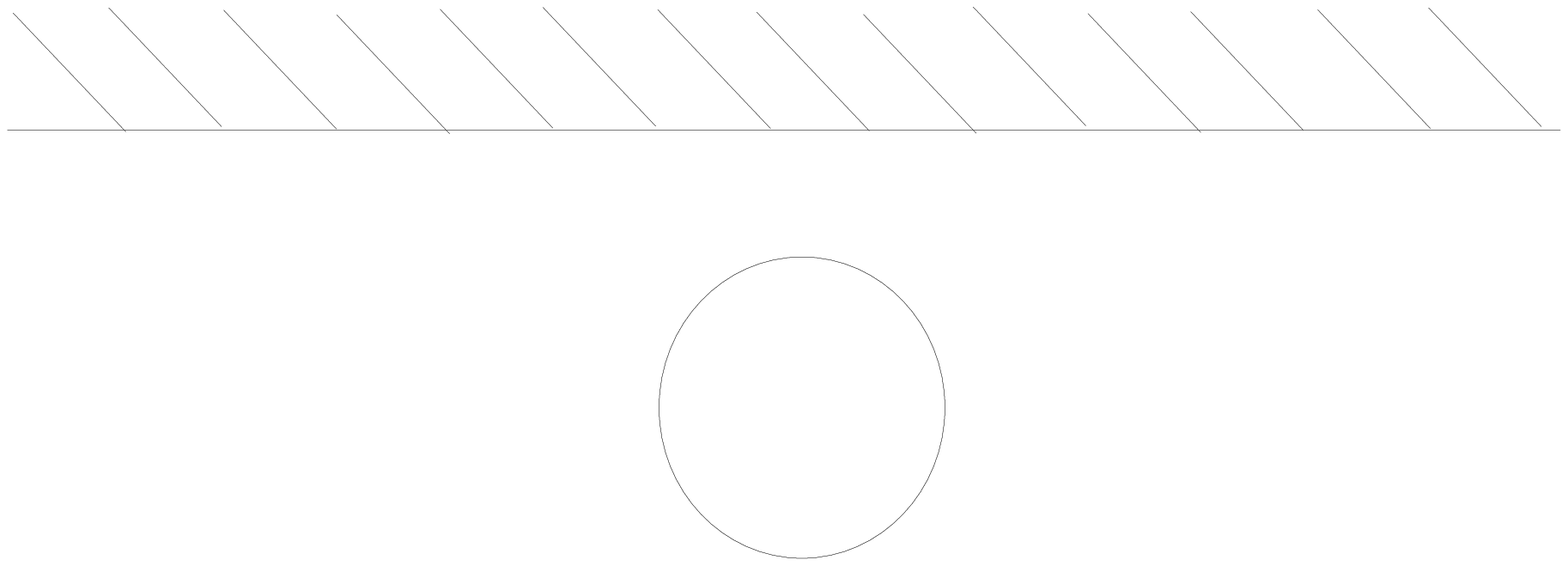}\hfil
\botcaption{Figure 2.}\endcaption
\endinsert
\pagebreak
\fi
Another configuration in which the horizon intersects the torus is
shown in fig 3
\iffigs
\midinsert
\hfil\leavevmode\epsfysize=7pc \epsfbox{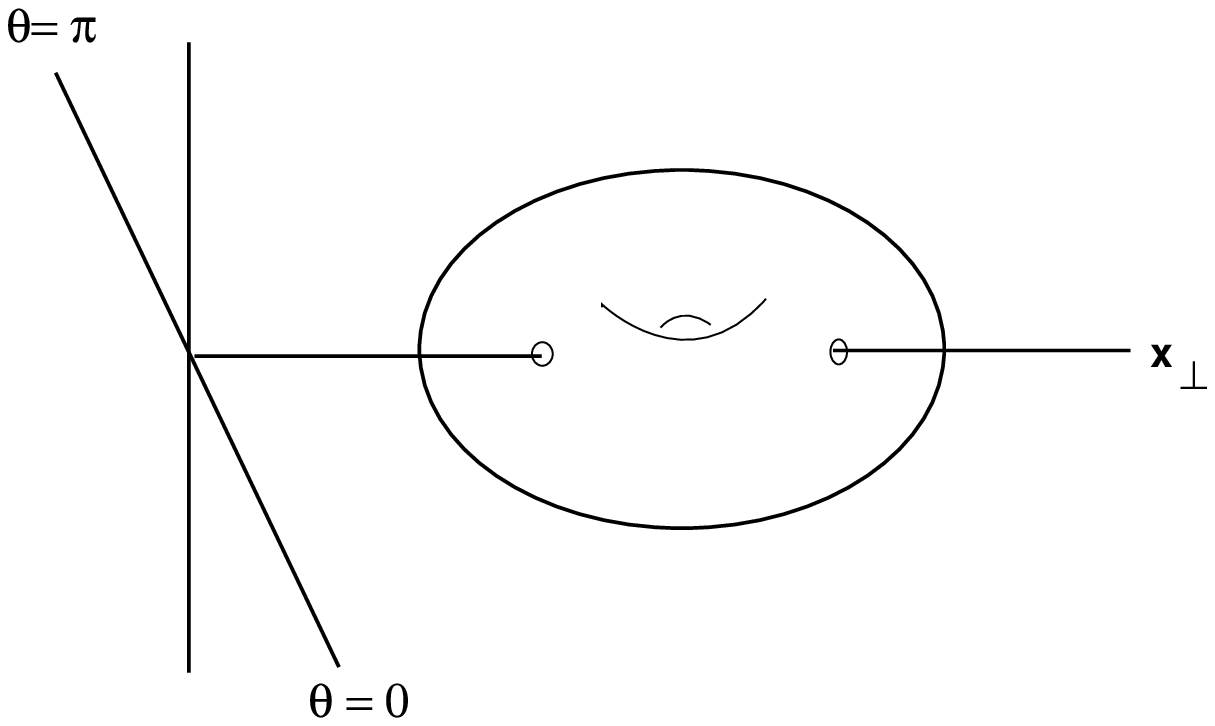}\hfil
\botcaption{Figure 3.}\endcaption
\endinsert
\medskip
\fi
In this case the instantaneous configurations involve strings with
their ends frozen on the horizon as in fig 4
\iffigs
\midinsert
\hfil\leavevmode\epsfysize=7pc \epsfbox{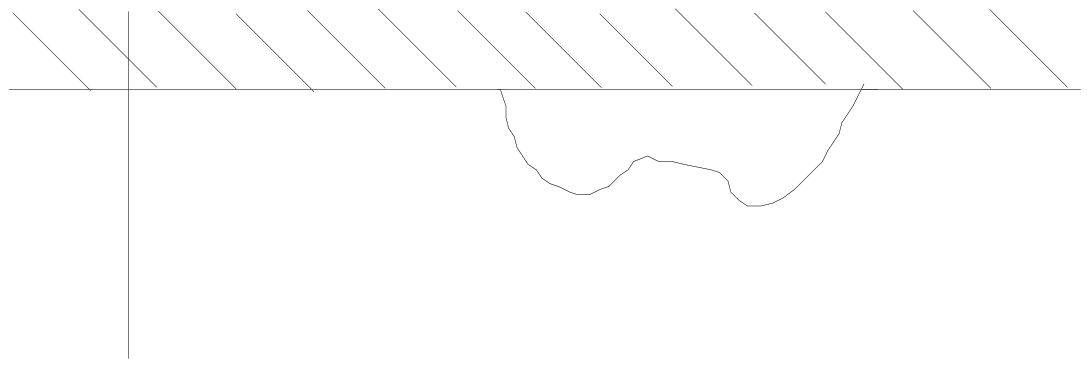}\hfil
\botcaption{Figure 4.}\endcaption
\endinsert
\medskip
\fi
Such configurations must be included in the space of states of the
black hole.
\medskip
\iffigs
\pagebreak
\fi
The genus $k$ surfaces contribute with a coefficient  $G^{k-1}$.
Evidently the Bekenstein entropy corresponds to a genus zero surface
with the topology of a sphere as in fig 5
\iffigs
\midinsert
\hfil\leavevmode\epsfysize=7pc \epsfbox{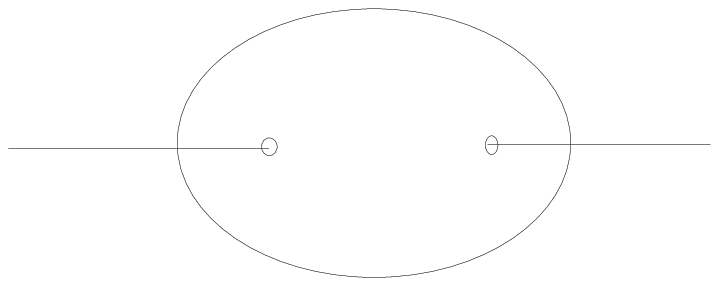}\hfil
\botcaption{Figure 5.}\endcaption
\endinsert
\medskip
\fi
These surfaces describe the evolution of a single string with ends on
the horizon.  In Minkowski space the endpoints of the string can not
move because of the infinite time dilation at the horizon but the rest
of the string is free to wiggle.  This solves the puzzle of the origin
of $S_B$ in String theory.
\medskip
One might wonder why the zero genus contributions do not vanish as
they usually do in string theory.  The reason involves both the
conical singularity at the origin and the fact that the deficit angle
at infinity does not vanish.  The presence of these genus zero
contributions very near the horizon is almost certainly related to a
similar effect found by Atick and Witten\cite{4} in high temperature
string theory and large $N$ $QCD$.
\medskip
We next turn to the question of the finiteness of the higher genus
quantum corrections.  In the absence of precise tools for quantizing
strings on singular spaces I can only quote some circumstantial
evidence for finiteness.  However, before doing so we must take care
of a trivial divergence associated with the infinite volume at spatial
infinity.  If the black hole is in thermal equilibrium with its
environment, the nonvanishing temperature will cause a volume
divergence in all extensive thermodynamical variables.  This can be
easily overcome by passing to the limit of infinite black hole mass.
The resulting geometry is flat Rindler space with a deficit angle.  In
Rindler space-times of dimension greater than two the thermodynamical
variables are infrared finite.  The entropy per unit area is therefore
well defined.
\medskip
As I have mentioned in sect 2 the ordinary field theoretic divergences
in entropy arise from paths of vanishing length which encircle the
horizon.  In string theory the analogue of a path of given proper time
is a torus with complex modular parameter $\tau$.  It is well known
that the integration region over $\tau$ which could potentially cause
ultraviolet divergences should be excised because it corresponds to an
infinite overcounting of geometrically identical tori.  There is no
mechanism for generating ultraviolet divergences if the relevant loop
integrals are anything like other string amplitudes.
\medskip
The finiteness of string theoretic loops is due to the extreme paucity
of degrees of freedom at short distances.  This lack of short distance
structure is seen in several ways.
\medskip
1)  Atick and Witten have shown\cite{4} that string theory behaves more
softly at high temperatures than any possible continuum quantum field
theory.  Roughly speaking the high temperature thermodynamics is
consistent with a lattice theory in which the spacing is finite.
Similar evidence comes from the work of Klebanov and  Susskind\cite{5}
who show that exact string amplitudes can be derived from a space-time
lattice theory with non vanishing spacing.
\medskip
2)  High energy scattering amplitudes at fixed angle are the
traditional method of uncovering short distance structure.  Gross and
Mende\cite{6} have shown that such amplitudes vanish like gaussian functions
of momentum transfer.
\medskip
3)  It appears to be impossible to force the dimensions of compact
space dimensions to be smaller than a certain size of order $\ell_s$.\cite{7}
\medskip
4)  Following the progress of a string falling toward a horizon, an
external observer fails to see the object Lorentz contract.\cite{2}  There
appears to be a minimum longitudinal size that strings can occupy.
Furthermore for non vanishing coupling there is a bound to the number
of strings that can pass through a small region without inducing
violent interactions.
\medskip
All these facts point to a common conclusion.  When we attempt to
localize strings or parts of strings in distances much smaller than
$\ell_s$ we discover a complete lack of local degrees of freedom.
This strongly suggests that higher genus contributions to black hole
entropy is finite and that to an external observer, indefinite
quantities of information can not collect arbitrarily near the event
horizon.  What is desperately needed is a computation to confirm this.
\bigskip
{\bf IV.  Black Holes as Single String States}
\medskip
I turn now to a radically different way to estimate the entropy of a
black hole.  When strings fall on to a horizon an external observer
sees them spread out and eventually fill the stretched horizon.\cite{2}
One can regard this phenomenon as a melting of strings as they
encounter Hagedorn temperature conditions\cite{4} at a distance
$\sim\ell_s$ from the event horizon.  The entropy of single string
states is so large that strings on the horizon will tend to form a
single string when the Hagedorn temperature is approached.  The
implication is that all black hole states are in one to one
correspondence with single string states.
\medskip
Now it has been observed in the past that the high mass-low angular
momentum states of string theory must be black holes since they lie
within their Schwarzshild radii.  I would like to make the
heretical suggestion that the spectrum of black holes and the spectrum
of single string states are identical.\cite{3}  Furthermore this provides
us with a direct way to estimate the number of levels and therefore
the entropy.
\medskip
Before we can actually compare the spectra we must deal with somewhat
trivial but numerically very important effect.  The classical
gravitational field outside the stretched horizon is not a low order
effect.  We must imagine removing the effects of this field before we
try to compare low order string theory with black hole physics.  The
main effect which must be accounted for is the large red shift of
clock rates that takes place between the stringy stretched horizon and
an observer at infinity.  In other words a rescaling of all energy
levels of the black hole should be done.
\medskip
The stretched horizon is the place where the local Unruh temperature
becomes hot enough for stringy effects to become important, i.e. the
Hagedom temperature.  This means a distance $\sim \ell_s =
(\alpha^1)^{-{1\over 2}}$ from the event horizon at $ r = 0$.  At this
distance, the proper time of a fiducial observer is given by
$$
\tau = \imath\theta\ell_s\tag4.1
$$
where $\imath\theta$ is Rindler time.  All quantities with units of
energy should be rescaled by dividing the corresponding Rindler
quantities by $\ell_s$.  The resulting quantities would also be
appropriate for an observer at infinity if the effects of the
classical gravitational field could be removed.
\medskip
Thus we define the stretched horizon energy and temperature of the
black hole to be
$$
\align
E(S.H.) &= {2 M^2 G\over \ell_s}\\
T(S.H.) &= 1/2 \pi \ell_s\tag 4.2
\endalign
$$
\medskip
Another way to think of the relation between $E(S.H.)$ and $M$ is that
the long range field outside the stretched horizon renormalizes the
mass from its ``bare'' value $E(SH)$ to its renormalized value $M$.
Therefore I would like to suggest that a black hole of mass $M$ should be
identified with a string state of mass $E(S.H.)$.
\medskip
We can estimate the entropy of a black hole by counting the levels
of fundamental strings.  The number of states at mass level $r = E$
satisfies
$$
\log N(E) \sim E\ell_s\tag4.3
$$
Using 4.2 we find a black hole of mass $M$ has a level density
satisfying
$$
\log N(M) \sim M^2G\tag 4.4
$$
which says that the entropy is of order $S_B$.  It is therefore not
inconsistent to suppose that a correspondence exists between black
holes and fundamental string states.
\medskip
Vafa has pointed out that this correspondence may extend to extreme
charged black holes.  In this case the mass renormalization due to the
gravitational field energy exactly cancels the electromagnetic energy
so that string states of mass $M$ should be compared with black holes
of mass $M$.  If, for example, we consider charge arising from winding
modes then the minimum mass of a string of charge $Q$ is proportional
to $Q$.  This corresponds exactly to the extremal black hole.
\medskip
If the view of black hole entropy in this section is correct then
there can be little doubt that the quantum corrections to $S_B$ are
finite.  These quantum corrections would result from the finite
shifting of levels due to higher genus world sheets.
\medskip
Finally I would like to mention an observable effect of the
string theory on black hole evaporation.  In the usual
picture the final evaporation process takes place at planckian
temperatures.  The last radiated particles would carry energy of order
the Planck mass.
\medskip
To best appreciate the difference that string theory makes, it is
helpful to pretend that the  string coupling $g^2 = G/\ell^2_s$ is
extremely small so that the Planck and string scales are extremely
well separated.	 Let us consider the radius of an average excited
string state of mass M, ignoring all higher order effects including
the long range gravitational field.  Ignoring quantum fluctuations of
the string the mean radius of the average configuration of mass M can
be shown to be
$$
R_{ST} = \sqrt{M\ell^3_s}\tag 4.5
$$
Comparing this with the Schwarzschild radius, $R_{SCH} \sim MG =
Mg^2\ell^2_s$, we see the two are equal when
$$
M = M_a = (\ell_sg^4)^{-1}\tag 4.6
$$
Thus for $g \ll 1$ there is a large range of masses for which the
conventional string configuration is larger than $R_{SCH}$ and no black
hole behavior should occur.
\medskip
On the otherhand an evaporating black hole should behave
conventionally until the red shift factor at the stretched horizon is
of order unity.  At this point there is no large red shift factor and
strings should behave like strings.  The mass at this point is
$$
M = M_b = {1 \over g^2\ell_s}\tag 4.7
$$
\medskip
Several things happen at this point.  The first is that the area of
the black hole horizon has become equal to $\ell_s^2$.  Second, the
Hawking temperature has reached the Hagedorn temperature.  Third, the
conventional mass M and the stretched horizon energy E(S.H.) become
equal.  Finally at this point the Bekenstein entropy $\sim M^2G^2$
becomes of order the string entropy $M\ell_s$.
\medskip
Between $M_a$ and $M_b$ we have two descriptions, ie. string and black
hole, both of which should apply but which disagree.
\medskip
The resolution of this inconsistency is that in this region there are
two configurations.  The first is metastable and describes a
conventional string with radius larger than $R_{SCH}$.  The second,
with larger entropy is stable and consists of a black hole with the
string gravitationally collapsed to the stretched horizon.
\medskip
Below mass $M_b$ the string entropy exceeds the black hole entropy so
that the black hole becomes thermodynamically unstable.  Accordingly
when a black hole reaches the Hagedorn temperature it ``inflates'' to
a string whose size exceeds that of a black hole by the factor
$$
{R_{ST} \over R_{SCH}} \sim {1 \over g}\tag 4.8
$$
Thereafter it decays like a weakly coupled string.  In particular the
momenta of the emitted particles never exceeds the Hagedorn
temperature.
\bigskip
\centerline{\bf Acknowledgments.}
This paper was written while the author was visiting
Rutgers University.  He is grateful not only for the hospitality, but
for many exciting discussions with the members of the theory group and
other visitors.  In particular C. Vata pointed out the significance of
extremal charged states in string theory.  The observations about the
final stages of evaporation arose in a conversation with Steve Shenker
about possible experimental signatures.  As always discussions with
Tom Banks, Nathan Seiberg and Mike Douglas were very valuable and fun.
Helpful discussions with Ed Witten and Curt Callan are also appreciated.
\bigskip
\bigskip
\bigskip
\bigskip


\bye